\documentclass{ieeeaccess}
\usepackage{cite}
\usepackage{amsmath,amssymb,amsfonts}
\usepackage{algorithmic}
\usepackage{graphicx}
\usepackage{textcomp}
\def\BibTeX{{\rm B\kern-.05em{\sc i\kern-.025em b}\kern-.08em
    T\kern-.1667em\lower.7ex\hbox{E}\kern-.125emX}}
\usepackage{dblfloatfix}
\usepackage{epstopdf}

\usepackage{url}

\usepackage{multirow}
\usepackage{comment}
\begin{document}
%
\doi{TBA}
\title{
End-to-end Ensemble-based Feature Selection for Paralinguistics Tasks}
\author{\uppercase{Tamás~Grósz}, Mittul~Singh, Sudarsana~Reddy~Kadiri, Hemant~Kathania, \\Mikko~Kurimo~\IEEEmembership{Senior Member,~IEEE}}
\address{Department of Signal Processing and Acoustics, Aalto University, Finland. (e-mail: firstname.lastname@aalto.fi)} 
\tfootnote{The authors would like to thank Antonia Hamilton and Alexis Macintyre for granting access to a subset of the UCL Speech BreathMonitoring (UCL-SBM) database used in the Breathing Sub-Challenge.
This work was supported by the Academy of Finland (grants 329267, 330139) and the Kone Foundation.  Aalto ScienceIT provided the computational resources.}

\corresp{Corresponding author: Tamás~Grósz (e-mail: tamas.grosz@aalto.fi).}

\markboth
{Grósz \headeretal: 
End-to-end Ensemble-based Feature Selection for Paralinguistics Tasks}%
{Grósz \headeretal: 
End-to-end Ensemble-based Feature Selection for Paralinguistics Tasks}%
%



\begin{abstract}
The events of recent years have highlighted the importance of telemedicine solutions which could potentially allow remote treatment and diagnosis. Relatedly,  Computational Paralinguistics, a unique subfield of Speech Processing, aims to extract information about the speaker and form a important part of telemedicine applications. In this work, we focus on two paralinguistic problems: mask detection and breathing state prediction. Solutions developed for these tasks could be invaluable and have the potential to help monitor and limit the spread of a virus like COVID-19. The current state-of-the-art methods proposed for these tasks are ensembles based on deep neural networks like ResNets in conjunction with feature engineering. Although these ensembles can achieve high accuracy, they also have a large footprint and require substantial computational power reducing portability to devices with limited resources. These drawbacks also mean that the previously proposed solutions are infeasible to be used in a telemedicine system due to their size and speed. On the other hand, employing \textit{lighter} feature-engineered systems can be laborious and add further complexity making them difficult to create a deployable system quickly. This work proposes an ensemble-based automatic feature selection method to enable the development of fast and memory-efficient systems. In particular, we propose an output-gradient-based method to discover essential features using large, well-performing ensembles before training a smaller one. In our experiments, we observed considerable (25-32\%) reductions in inference times using neural network ensembles based on output-gradient-based features. Our method offers a simple way to increase the speed of the system and enable real-time usage while maintaining competitive results with larger-footprint ensemble using all spectral features. 
\end{abstract}


\titlepgskip=-15pt

\maketitle

\IEEEdisplaynontitleabstractindextext

%
\IEEEpeerreviewmaketitle

\IEEEraisesectionheading{\section{Introduction}\label{sec:introduction}}

\IEEEPARstart{E}{v}ERY year INTERSPEECH's computational paralinguistic challenge (ComParE)\footnote{http://www.compare.openaudio.eu/} has pushed the boundaries of speech signal processing by showcasing interesting speech-related prediction tasks. This interdisciplinary research area, which focuses on extracting information from speech beyond what the speaker said, has helped promote solutions for telemedicine systems~\cite{9191596,Schuller2020COVID19AC} during the time of COVID-19 pandemic.
 
In this work, we consider two computational paralinguistic tasks that were part of the INTERSPEECH 2020 ComParE challenge: 1) \textbf{Mask Detection} -- to predict if the speaker wears a mask or not while they speak, and 2) \textbf{Breathing State Detection} -- to predict the output signal of a respiratory belt worn by the speaker. Both mask and breathing state detection tasks are relevant to combat the COVID-19 pandemic~\cite{Schuller2020COVID19AC}. 

Due to COVID-19, mask-wearing has become a widespread habit of the general population, and studies show that masked speech requires special care as mask could cause reduced inteligibility~\cite{10.3389/fpsyg.2021.682677}. This means that mask detection solutions can help with COVID-19 risk assessment of individuals~\cite{Schuller2020COVID19AC}, as well as a valuable part of speech recognition systems informing the model about the speaking conditions. The second task, breathing state detection, exemplifies the possibilities of remote patient monitoring and diagnosis.

State-of-the-art models for such prediction tasks, pay little attention to latency. They often utilize ensembles of neural networks, which have a large memory footprint and are slow for on-device usage. In this work, we take a step towards designing systems that balance accuracy and speed. To embed the tools in smartphones, we need to be efficient with resource utilization, especially with on-device memory usage and inference time to provide results quickly and efficiently. 

Starting with a large ensemble of end-to-end neural models, we have two options to create an equally efficient but smaller and faster solution. The first option is knowledge distillation, which aims to transfer the learnt knowledge from the large ensemble to the smaller models~\cite{asif2019ensemble}. The second option is feature selection. In our work, we opted for the second one, which simultaneously reduces the size of the models and the input data. This leads to better latency and lowers network usage in the case of a server-client type of system. We propose a novel way of feature selection with ensemble neural networks to reduce memory usage and increase speed, thus making the neural ensembles a viable solution. Our feature selection method considers output gradients of ensemble neural networks to filter essential features for further model development. 

On mask and breathing state detection, the output-gradient-based models achieve competitive performance with the baseline and it is able to reduce the memory usage and inference time significantly. Furthermore, in comparison to state-of-the-art methods, the output-gradient based models provide substantial reduction in memory and inference time while maintaining a good performance.

\subsection{Prior works on INTERSPEECH 2020 Computational Paralinguistics Challenge}
The mask detection and breathing state detection were studied in the sub-challenges of the INTERSPEECH 2020 ComParE challenge. 
For the mask detection sub-challenge, authors in \cite{Yang2020} explored the  Fisher Vector (FV) encoding over the traditional mel-frequency cepstral coefficient (MFCC) features and found promising results. FVs also found to provide complementary information when fused with other acoustic features. The authors also investigated several fusion techniques; Majority Vote, Harmonic Mean, support-vector machine (SVM) fusion,
and deep neural network (DNN) fusion to combine various acoustic information from baseline features such as ComParE functionals, bag of audio words (BoAWs), Deep Spectrum and auDeep~\cite{Freitag2017AuDeep}, and demonstrated the contribution of FV-MFCC feature in model fusion. 

Authors in \cite{Montacie2020} have investigated the phonetic context and intelligibility measurements to determine whether a speech signal produced with or without wearing a surgical mask. The most sensitive phonemes are nasals, front and back vowels for the speech produced with mask wearing. Two systems were developed, one using a larger audio feature set and the second using a bottom-up approach with phonetic analysis and clustering. Fusion of the systems were shown to improve the performance over the baseline. In \cite{Klumpp2020}, authors proposed the usage of deep recurrent phoneme recognition model which was trained on spectrograms to learn the spectral properties of various speech sound categories. Their assumption is that, each phoneme sounds differently between mask and clear speech. The phoneme recognition model consists of a convolutional network for feature extraction from the spectrogram and a recurrent network for sequential phoneme classification. The phoneme recognition model was used to differentiate between mask and clear phoneme productions. Phonetic representations were shown to perform better compared to well-established acoustic features as well as deep feature representation in mask speech detection. 

In another work \cite{Szep2020} the use of pre-trained deep convolutional neural network-based image classifiers is investigated with spectrograms. In it, specifically three different image classifier architectures were used: VGGNet, ResNet, and DenseNet. Several types of spectrograms were investigated for input, such as linear and quasi-logarithmic frequency-scale, Mel-spectrograms, Constant-Q Spectrograms, and MFCCs. The ensemble of models found to improve the accuracy for mask challenge.

Perhaps the most related work from the literature is \cite{Markitantov2020}, in which end-to-end deep Convolutional Neural Network (CNN) architectures are explored with raw waveforms, log Mel-spectrograms, and MFCCs. For mask sub-challenge, experiments were conducted with several pre-trained CNN architectures on log-Mel spectrograms, as well as SVMs on baseline features. For the breathing state detection sub-challenge, an ensemble deep learning system that exploits end-to-end learning and sequence prediction were investigated. Also, authors have shown that by effectively increasing the training data via N-Fold cross-validation and careful selection of pretrained models, performance can be improved. 

In \cite{MacIntyre2020}, learning the speech features in an end-to-end architecture was shown to be beneficial for breathing sub-challenge, and also the attention mechanism was shown to be helpful in identifying the useful patterns. The information present in the AM-FM (amplitude modulation-frequency modulation) decomposition was shown to be helpful in characterizing the \textit{breathiness} from the speech signal in \cite{Mendonca2020}. For AM-FM decomposition, frequency domain linear prediction (FDLP) approach is used. End-to-end architecture (similar to baseline in \cite{compare2020}) was used for the experiments. Authors also analysed the original and predicted signals to overcome the main pitfalls, and found that the subsets of most irregular belt signals yielded the worst results (measured by the Pearson correlation coefficient). This indicates the effect on results that were obtained by baseline systems and various proposed systems.

Studies in \cite{Illium2020,Koike2020,Ristea2020} explored the data augmentation  to overcome the lack of training data for deep neural architectures for mask speech detection. Specifically, five different augmentation methods were used in \cite{Illium2020}, they are: speed perturbation, loudness perturbation, shift perturbation, spec-augmentation, and noise augmentation. Transfer learning strategy combined with data augmentation was investigated for mask sub-challenge in \cite{Koike2020}. Instead of pre-trained models obtained from image data, authors used a pre-trained model developed on large amount audio data. Apart from pre-trained model, the data augmentation (spec-Augmentation and mixup (to increase volume) methods) and late fusion strategy were used to improve the performance. Novel  data augmentation method was proposed by transferring data of one class to another through cycle-consistent Generative Adversarial Networks (cycle-GANs) in \cite{Ristea2020} for mask speech detection. In this, the spectrograms of original and augmented data are used as input to set of ResNet neural networks with various depths and then the networks are combined into an ensemble through a SVM classifier.

\subsection{Highlights/contributions of this study}
\begin{itemize}
    \item E2E Ensemble solution
    \item New gradient-based feature selection method
    \item Exploration of new features for mask challenge
    \item Model fusion experiments 
\end{itemize}
As we can see, many works have found that E2E ensemble-based solutions are the best for these paralinguistic tasks. Unfortunately, one vital aspect is ignored by these works, namely the practicality of the systems. Using ensembles of multiple large models could eventually lead to extremely slow solutions that cannot be used by real-life applications. One key difference between our work and the previously published works is that we aim to use small, feasible models, which could become part of real-world applications. Furthermore, we also explore several input features and a new feature selection method that can pick the most important features using several well-performing DNNs.  These feature selection methods aim to reduce model sizes and inference times while maintaining a good performance.  Although our ensembles could not beat the best systems in the competition, they came quite close, while being much faster and smaller than the ensembles of the  state-of-the-art ResNet models.


\section{Computational Paralinguistics Sub-Challenges}
First, let us take a look at the data available for each of the sub-challenges. Since we used a deep learning-based solution, the quality and quantity of the training data are vital for success. The corpora are described in more detail in the baseline paper~\cite{compare2020}, here we focus on the most relevant properties of each corpus.

\subsection{Mask sub-challenge}
The Mask sub-challenge (MSC) focuses on detecting whether the speaker is wearing a mask or not.  
This task is crucial these days as the COVID-19 pandemic rages on in the world. Mask-wearing is recommended in many countries to slow the spread of the virus. The systems developed for this competition could be monitor if oneself or others around are wearing a mask when speaking~\cite{Schuller2020COVID19AC}.

What makes this extremely challenging is that the decision must be made using only 1 second of speech (typically one word). The audio data comes from the Mask Augsburg Speech Corpus (MASC), where 32 German speakers were recorded (16 male and 16 female). The total duration of the recordings is about 10 hours, which was split into three sets. The speakers were performing various tasks while speaking with or without a mask. Their activities include question answering, reading words from the professional medical vocabulary and describing pictures. 
The data was mostly balanced; approximately 5.5K samples were given for the masked category and 5.3K for the non-masked in the training data, while the development set contained almost 8K masked and 6.6K non-masked recordings. The test set did not differ from this distribution and consisted of about 5.5K recordings for each class.

\subsection{Breathing sub-challenge}
For the Breathing sub-challenge (BSC),  a subset of the UCL Speech Breath Monitoring (UCL-SBM) database~\cite{compare2020} was used. The speakers wore respiratory belts while speaking, and the database consists of the synchronized speech and belt signals. The aim here is to estimate the measurements of the belt from the recorded speech as closely as possible. 
Naturally, a system that can predict respiratory signals from speech without a belt has high potentials in real-life applications. Such algorithms could be run on smart devices and used to detect the symptoms of diseases like COVID-19~\cite{Schuller2020COVID19AC}, without visiting medical experts. As such, they could be essential tools to recognize the early signs of sickness and even aid diagnosis.

In total, 5 minutes of spontaneous speech was available from 49 speakers. All speakers reported English as their native language, with varying accents (e.g. American, Irish, etc.). The recordings contain spontaneous speech, all participants answered some questions about their experience of visiting or living in the city of London, and they were allowed to speak about different topics. This data was split into three subsets, 17 subjects were allocated to the training part, 16 for the development set, and the rest of the recordings were dedicated for testing purposes. This task is a so-called sequence-to-sequence problem, where using the audio, we need to predict the 6000 measurements produced by the respiratory belt while the patient spoke. The main challenge to overcome in BSC was the limited amount of data (only 17 training instances).



\section{Methods}
In this section, we describe the usage of end-to-end systems in an ensemble learning scheme as it is the core of our solution for the challenges. Other, task-specific techniques and modifications of the generic framework will be described in detail in Section~\ref{sec:exp_res}.

\subsection{End-to-end learning}
End-to-end learning is a new emerging paradigm within deep learning. Researchers across various fields have adopted this paradigm supported by the availability of large data and powerful computational resources. Theoretically,  end-to-end systems are built to replace the traditional pipeline-based solutions with a single deep neural network. The end-to-end systems allow using a single optimization step to training the complete model. They also have the promise of bypassing the laborious feature-engineering step by having a single system for solving every aspect of the prediction problem. In practice, however, these systems are built on top of existing features. The advantages of this paradigm make it an attractive choice for ComParE tasks. 

Our models process either spectral input features or raw audio signals in case of the breathing task. Then the DNNs can directly be optimized to perform the given task. Using this single model approach allowed us to quickly modify the general framework to the specialties of the sub-challenges.

\subsection{Ensemble learning}
DNNs are known to be sensitive to the random initialization, and our experiments also confirm this. This issue is especially severe if the amount of training data is limited, which is usually the case for paralinguistic tasks. A solution to this problem is applying ensemble learning. We train several differently initialized DNNs and then combine their predictions to get stable and even better results. 

Here, we employ a specific bootstrap aggregation method, called bagging. Originally, bagging trains each model using only a random subset of the training data to produce diverse systems. As the training data is already limited, we decide to use all available data during training and rely on random initialization and data shuffling to produce a diverse set of DNNs. In the combination, we average the outputs of differently-initialized DNN ($f_i i \in \{1..N\}$) on input $x$ together to make the final prediction ($\bar f$). For regression-based breathing task, the final prediction can be calculated as:
\begin{equation}
    \bar f(x) = \frac{1}{N} \cdot \sum_{i=1}^N f_i(x)
\end{equation}

For classification-based mask task, we apply a soft voting scheme to average output probabilities (prob$_i$) for every DNN $f_i$ obtain the final prediction. In this scheme, the ensemble probability ($\overline{\mathrm{prob}}$) of class $C$ for a given input $x$ can be obtained as follows:
\begin{equation}
    \overline{\mathrm{prob}}(C | x) = \frac{1}{N} \cdot (\sum_{i=1}^N\mathrm{prob}_i(C|x)
\end{equation}

The ensemble learning can also be performed with other approaches.
By combining different approaches we can leverage their individual strengths and create a better solution. 

\subsection{Input sensitivity analysis}

Traditionally, DNNs are viewed as black boxes, and during their evaluation, most studies only focus on the performance measures. This trend started changing as recent attempts have been made at investigating how and why DNNs can achieve high performance~\cite{Nguyen2019}. A convenient tool to examine trained DNN models is Sensitivity Analysis that aims to identify the crucial features using a trained DNN. The prerequisites for the analysis are only a well-trained model and some data. Of course, several ways are available to perform the sensitivity analysis like Saliency maps~\cite{simonyan2014deep}, Gradient-weighted Class Activation Mapping~\cite{selvaraju2017grad}, Layer-Wise Relevance Propagation~\cite{montavon2019layer}. Out of those, we focus on the Saliency maps-based approach, which was originally proposed to visualize how image processing DNNs function. Saliency maps are simply constructed by calculating each pixel's importance scores, gaining insights into which parts of the image have the largest influence on the model's output. 
In our work, we take this one step further and propose that Saliency maps can also be used for selecting important features.

\subsection{Input sensitivity-based feature selection}
\label{sec:feat_selection}
In our work, we aimed to understand the individual DNNs better and leverage the findings to build smaller and faster models. Calculating the Saliency maps~\cite{simonyan2014deep} gave us insights into which parts of the input was most influential. Our working hypothesis was that we could use this information to select the essential features to reduce the input size and make the training process more straightforward.
To calculate the individual features' importance, we calculate and aggregate the model's locally evaluated gradient w.r.t. the input neurons.

Naturally, the gradient calculation can be done in various ways. One such method called Activation Potential analysis~\cite{Roy2015DNNfeatSelection} proposes that the first hidden layers' activations can be used. This method assigns importance to the input features by determining their contribution to the hidden neurons activation potential. 
Another solution is to use the output of the DNN directly. In this case, we can choose to propagate either the output values or the loss values to find essential inputs. Other methods use more complicated approaches to calculate the importance score of inputs.  DeepLIFT~\cite{shrikumar17DeepLIFT} proposes that each neuron's activation can be compared to a task dependent 'reference value', then the contribution scores are determined by backpropagating these differences to the input layer.
 Another method, Integrated Gradients~\cite{sundararajan2016Intgradients} scales the inputs from some starting value to their actual value and integrates the gradients to calculate the final importance scores.

In our experiments, we employed two simple approaches; the output and loss-based gradients were compared to measure the individual features' impact on the whole network. This way, we can inspect the DNNs function and reduce the input at the same time.
Formally, let $x_i$ denote the $i^{th}$ input feature, and $\hat{y}=DNN(x)$ is the output value produced by the DNN. If we have the labels ($y$) available, then the loss function ($F_{loss}$) can also be calculated.
The importance then can be calculated using the output as:
\begin{equation}
    R(x_i) = \sum_{x \in D} |\frac{\partial\hat{y}}{\partial x_i}| ,
\end{equation}
or by propagating the loss values:
\begin{equation}
    R(x_i) = \sum_{x \in D} |\frac{\partial F_{loss}(y, \hat{y})}{\partial x_i}|.
\end{equation}
Lastly, to get the reduced feature set, we select the ones with the largest accumulated gradients.

Comparing these two methods, the first thing that we can notice is that the output-based one does not require labeled data. It can be applied to any data, technically making it an unsupervised method. In contrast, the loss-based selection can only be employed if we have ground truth labels. Another vital thing to note here is the fact that the two methods highlights different groups of features. While the output-gradient propagation ensures that we get the features that have the most significant impact on the network's decision, the loss-based one determines the problematic inputs responsible for the network's errors. The reason for this is simple; well-predicted examples will have low loss values, and thus the gradients propagated to the inputs will also be minimal. On the other hand, when the DNN makes mistakes, we get high loss gradients. Consequently, the propagation method will assign high importance to features responsible for the loss (i.e., inputs that can potentially change the loss value quickly).

\begin{figure*}[ht]
    \centering
    \includegraphics[width=\linewidth,height=11cm]{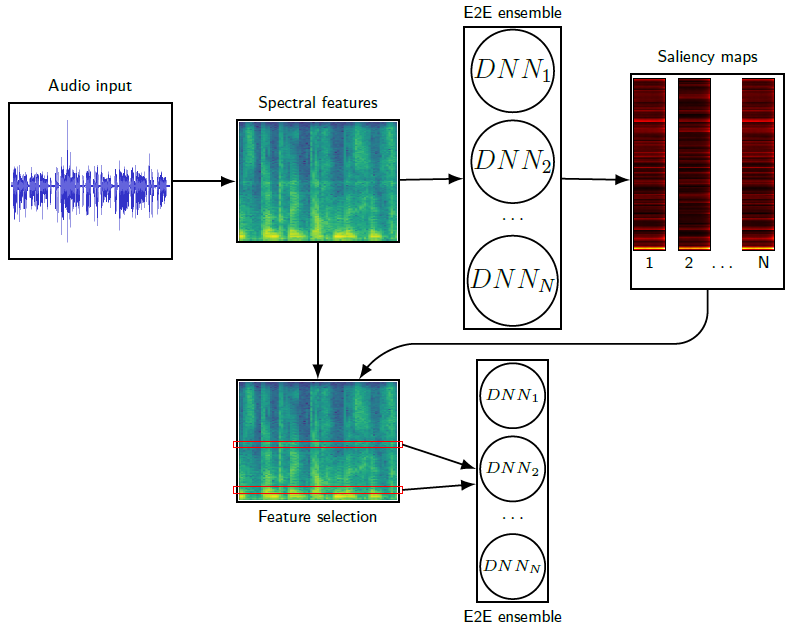}
    \caption{General workflow of our proposed input importance-based feature selection system. The saliency maps are computed using the outputs of an ensemble of large DNNs trained on the full feature set. The input importance scores helps us to select features on which we can train a new ensemble of smaller but more focused DNNs.}
    \label{fig:workflow}
\end{figure*}

The previously introduced techniques can be readily used with a single DNN, and here we show a simple way to apply them with ensembles of DNNs. We argue that relying on the features selected by a single model is dangerous as the selection could be a result of overfitting or some initialization anomaly. To avoid depending on single models, we propose a way to reduce the feature set using multiple networks. The procedure is simple; individual models select their $N$ most important features, and each DNN votes for it's chosen inputs. We then aggregate these feature sets by a majority voting scheme to select the $N$ most supported ones. This way, we can ensure that we use only the inputs supported by multiple models, reducing the risk of choosing features due to the previously mentioned problems. Our proposed workflow is depicted in Fig.~\ref{fig:workflow}.

\section{Experiments and results}
\label{sec:exp_res}
Next, we explain the experiments carried out for each task.
Keeping in mind that this paper describes our solutions for a competition, so we broke with the tradition of using only a limited amount of techniques. Instead, we tried several techniques to create the best system that can be deployed to hand-held devices. Keeping it in mind, we did our best to measure the impact of each modification on the development data and tested only the best ones. 

\subsection{Mask recognition}
For the MSC, we employed a relatively simple E2E DNN architecture. The spectral input is first processed by two 1D convolutional layers each containing 64 neurons. Afterward, a recurrent layer with 100 LSTM cells accumulates the outputs of the convolutional filters. We further process the outputs of the recurrent layer using a feedforward layer (100 rectified-linear units) before the classification layer. Our system was implemented using the Keras framework~\cite{chollet2015keras}, and TensorFlow~\cite{tensorflow2015-whitepaper} as the backend. 
During training, we utilized an Adam optimizer with a learning rate of 0.001. Using the development portion of the data, we found that 10 epochs of training yields the best results with a batch size of 100\footnote{Further details can be found in our repository: \url{https://github.com/aalto-speech/Compare2020}}.

\begin{table}[t]
    \centering
    \caption{Initial results on the Mask Sub-Challenge (MSC) for Development (Dev) set.}
    \begin{tabular}{c|c}
         System &  Dev \\ \hline
         Single best SVM~\cite{compare2020}& 64.4 \\
         E2E DNN & 66.0$\pm$4.8 \\
         Ensemble E2E & {\bf 67.4} \\ \hline
    \end{tabular}

    \label{tab:mask_init}
\end{table}

In table~\ref{tab:mask_init}, we show the initial results got by training an ensemble of 10 models. We can see that the accuracy of the ensemble is significantly better than the average performance of the individual E2E models or the best non-fused SVM system (DeepSpectrum+SVM)~\cite{compare2020}. The ensemble's DNN components achieved 66.0\% UAR on average, with high variation ($\pm4.8$), showcasing that DNNs are quite sensitive to random initialization, especially when the amount of training data is limited. To validate that the models managed to capture different aspects of the data and are suitable for building an ensemble, we also calculated the inter-model agreements, which varied between 69\% and 89\%, with an average of 81.5\%. The combination of the 10 models produced 67.4\% UAR, which is markedly better than the average performance of its individual components. From this, we can safely assume that ensembling E2E models is indeed beneficial.

Although our E2E ensemble outperformed the best non-fused systems in~\cite{compare2020}, which uses auDeep features and an SVM model, the fused solution presented in the competition paper still proved to be better than our approach. This superior system consisted of several SVMs trained on various features like auDeep-fused, Bag-of-audio-word, OpenSmile, and DeepSpectrum features~\cite{compare2020}, which means that it is considerably slower than our solution; the DeepSpectrum component alone requires several order of magnitude more time than our E2E ensemble (see Table~\ref{tab:res_select}).  Nevertheless, the observations that combining simple SVM models trained on different input features could improve the performance prompted us to investigate other input features for our E2E ensemble.

\begin{figure}
    \centering
    \includegraphics[trim={0 0 1.5cm 0},clip, height=7cm]{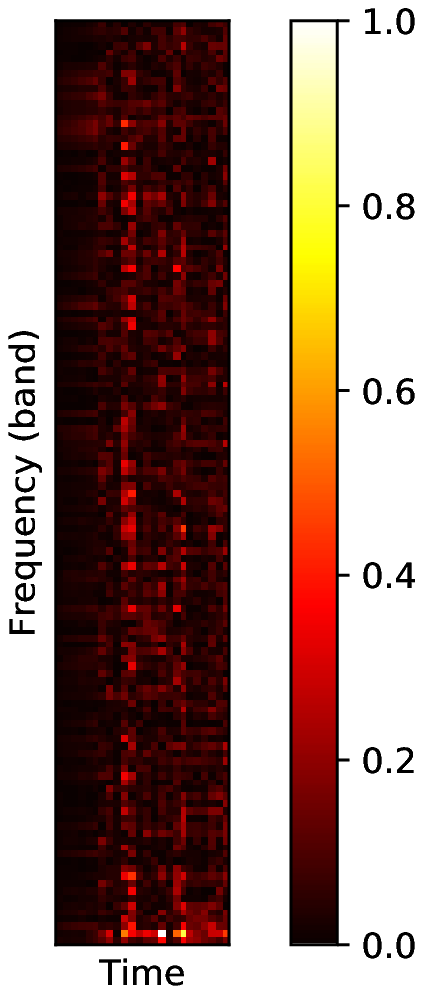}
    \includegraphics[trim={0.5cm 0 1.5cm 0},clip,height=7cm]{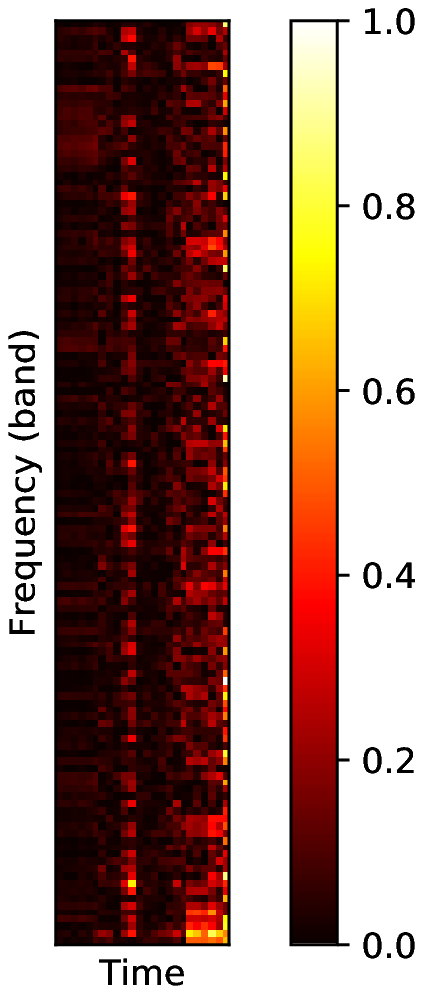}
    \includegraphics[trim={0.5cm 0 1.5cm 0},clip,height=7cm]{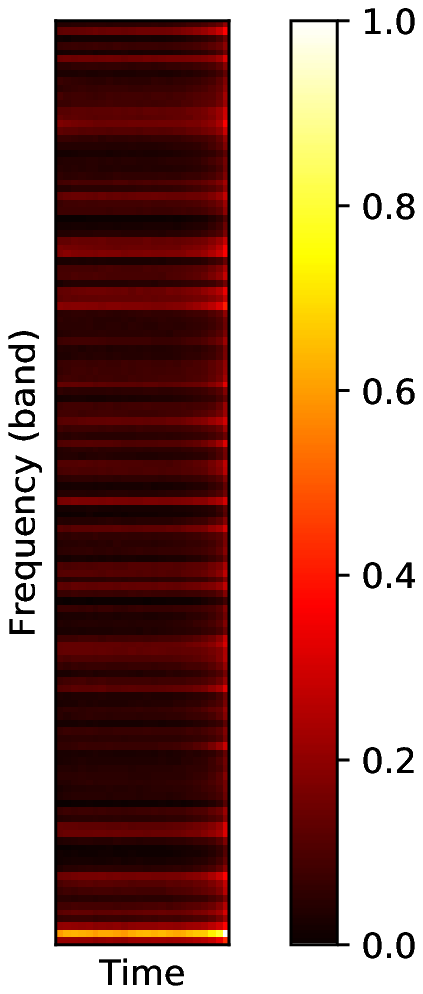}
    \includegraphics[trim={0.5cm 0 0 0},clip,height=7cm]{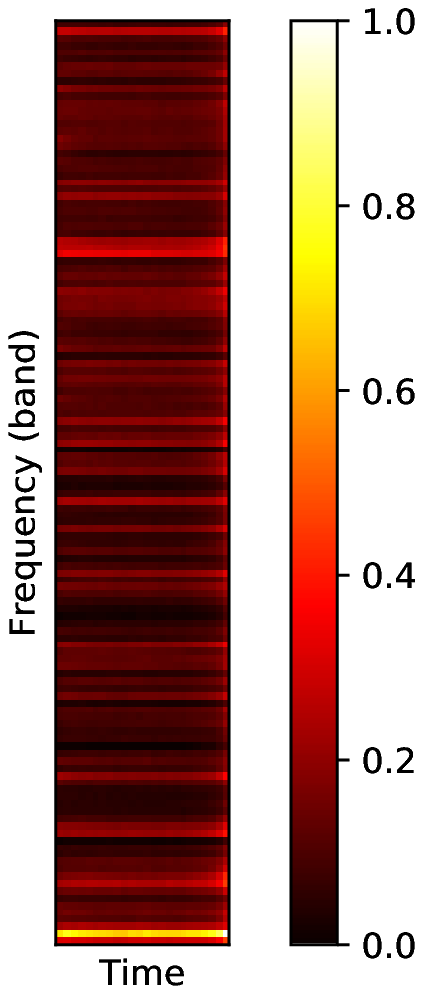}
    \caption{The figure shows the gradient magnitudes of DNN outputs with respect to the input spectral features on the Mask sub-challenge. The left two images show the gradients of the two randomly selected models for a single audio file and the right two images show the gradients of the same two models averaged over all training files. A bright yellow shade represents the largest gradient magnitudes seen for the lowest frequencies.}
    \label{fig:low-freq-grads}
    \vspace{-0.5cm}
\end{figure}

To further improve our system, first, we examine the hypotheses that wearing a mask changes the resonance conditions in the vocal tract, as the mask might reflect some of the frequencies to the tract~\cite{phonation_speech,Mel_phonation}. To confirm the hypothesis we first inspected our models. Since our solution uses E2E models, it is relatively easy to inspect them to determine their most important input features. 
Here, we employed input sensitivity analysis using the training data and plotted input saliency maps in Figure~\ref{fig:low-freq-grads}. Noticeably, the bands in the low-frequency domains have high importance scores, meaning that this part of the spectrum has the most influence over the output. This observation inspired us to design and test new features that focus on the signal's low-frequency elements.

\subsubsection{Low-frequency features}
To enrich our feature set, we tested two straightforward approaches. The first solution utilizes a pre-processing step, enhancing the lower frequencies by manipulating the input audio. We apply low-frequency enhancing schemes like pre-emphasing the audio (with filter coefficients $h=[1~0.97]$ and passing through a fifth-order low-pass Butterworth filter whose cutoff frequency is $400~Hz$~\cite{oppenheimBook}, before spectral feature extraction, denoted as \textbf{preemphasized spectrogram}. This process could allow the Mel-spectrogram to represent the relevant information for this task better if it is hidden in the low-frequency bands. 

The second approach is even simpler, it represents the information in a given time frame by calculating the ratio between the energies in the low- and high-frequency domain (denoted as Freq. ratio).

In Table~\ref{tab:new_feats}, we can see that models trained on these new features failed to outperform the ensemble utilizing the original spectrogram. Nevertheless, we kept these features, hoping that we can use them to build a better system by fusing them to the original input. Ultimately, these experiments showed us that the low-frequency part is not the only vital input region.

\begin{table}[h]
    \centering
    \caption{Results got by using our custom (low-frequency) features on the Dev set of the Mask Sub-Challenge.}
    \begin{tabular}{c|c}
         Features & Dev \\ \hline
         Original spectrogram & 67.4\\
         Pre-emphasized spectrogram & 59.3\\
         Freq. ratio & 57.2 \\\hline
    \end{tabular}

    \label{tab:new_feats}
\end{table}

\subsubsection{Feature selection}
\begin{table}[t]
    \centering
    \caption{Results of the feature selections approaches for the Mask Sub-Challenge. Inference time calculated as the average time required to evaluate the ensemble of 10 DNNs on one example (including feature extraction) from the Dev set using a single core of an Intel Xeon W-2133 CPU.}
    \begin{tabular}{c|c|c|c}
        Features & Dev & Inference time & Model size\\ \hline
        DeepSpectrum~\cite{compare2020} & 63.4 & 3090 ms & 23M\\\hline
        Complete spectrogram & 67.4 & 4.1 ms & 88K\\\hline 
        Lowest 10 bands & 62.9 & \multirow{6}{*}{2.8 ms} & \multirow{6}{*}{80K}\\ 
        10 least important (output) & 59.7 &&\\
        10 random & 62.0 &&\\
        10 most important (loss) & 65.1 &&\\ 
        10 most important (SFFS) & 65.9 && \\
        10 most important (output) & \textbf{67.9} &&\\\hline
    \end{tabular}

    \label{tab:res_select}
\end{table}

A natural choice for this is to select the ones deemed important by the initial models. We tested multiple different approaches. As a simple baseline, we tested using only the 10 lowest frequency bands. To determine the frequencies with the highest impact, we calculated the importance scores by attributing the output or the loss of the model to the inputs (as described in Sec.~\ref{sec:feat_selection}). All models in the initial ensemble voted for the indices of their 10 most valued bands. After that, we selected the ones that received the most vote. The choice to use only 10 bands was based on the fact that these features had majority support, meaning that at least 5 out of the 10 models voted for them. To validate that the importance scores are indeed meaningful, we also trained two systems, one with the least important features and one with 10 randomly selected frequency bands. Table~\ref{tab:res_select} confirms that using only a small portion of the original features can lead to good results. It is also clear that the naive approach to use only the low-frequency part (Lowest 10 bands) is inferior to the feature selection systems; still, it managed to outperform our handcrafted features. Comparing the two selection criteria, we saw the output gradient-based importance scores lead to better models. This ensemble even managed to outperform the one that was trained using the whole spectrogram. Furthermore, choosing the 10 least important features led to the worst results, while the random selection performed slightly worse than the naive baseline. 

To validate that our feature selection solution is well suited for E2E solutions, we compared it with the widely used Sequential Forward Feature Selection (SFFS) algorithm~\cite{FERRI1994403}. The SFFS algorithm starts by selecting the single best feature and then sequentially adds features that produce the best results. Its main drawback is that it requires the training of a large number of models, to be precise, we needed to train 1235 models to select the 10 best frequency bands out of the 128. Naturally, using the E2E models in this procedure is not feasible due to their considerable training time. Instead, we opted to use simple linear SVMs as proxies during the selection process and then trained the E2E ensemble using the chosen features. In Table~\ref{tab:res_select}, we can see that SFFS yielded better results than our loss-based feature selection, but it failed to reach the performance of the output-based selection method.

Another important aspect we need to consider is the speed and model size. The networks that used the selected features were significantly faster than those that used all 128 frequency bins. They had 10\% fewer parameters. During inference, they operated on far less data (10 features vs 128 frequency bands). Please note that our inference time measurement also includes the time needed for spectrogram extraction (on average 1 ms per example).  Overall, the reduced models needed about half the time (inference time reduced by 45\%) to process the inputs compared to the ensemble working with the complete spectrogram.

\subsubsection{Feature fusion}
At this stage, we had a wide range of features at our disposal. In the challenge paper \cite{compare2020}, we saw that fusing different features is quite beneficial. This inspired us to investigate whether training models using a granularity of features could offer any improvements. 

After some initial experiments, we decided to use the middle fusion technique. This meant that our DNNs first processed each input separately and only concatenated the outputs of the LSTM layer before the feedforward layers. Having a dedicated convolutional and recurrent part for each feature allows the network to extract meaningful latent information from each modality. The joint feedforward layer's task in this setup is to combine the different feature-dependent pieces of information for the final classification. This approach proved to better than using early- or late fusion.

Since we had to test several combinations to select the best, we decided to narrow down the number of choices. Our preliminary experiments established that it is always beneficial to show the original spectrogram data to the models, so we tried fusing the other features with the original spectrogram. In the end, we opted to try the 10 features selected using the output gradient-based importance scores, the frequency ratios and the pre-emphasized spectrograms.

Table~\ref{tab:fusion_res} reports the results of the input fusion experiments. As we can see, fusing the 10 most influential bands with the original data lead to the best results, meaning that our system managed to combine the knowledge extracted from these redundant sets. A possible explanation for this is that the network part that worked on the selected features could extract the most relevant information; simultaneously, the part that operated on the whole spectrogram could focus on the rest of the bands to extract extra information. The pre-emphasis and frequency ratio features on their own were not useful, but fusing both of them to the original spectral input, we saw some improvements. This shows that it is indeed important to utilize a sufficiently rich feature representation of the task.

\begin{table}[h]
    \centering
    \caption{Results got by middle fusion of various input types using ensembles of 10 models for Mask Sub-Challenge. '+': inclusion and '-':exclusion}
    \begin{tabular}{c|c|c|c||c}
        \multicolumn{4}{c||}{Inputs} & \multirow{2}{*}{Dev}  \\ 
        orig. spect. & selected 10 & pre-emph. & freq. ratio &\\ \hline
         + & -- & -- & -- & 67.4  \\ \hline
        + & + & -- & -- & 68.4  \\
        + & -- & + & -- & 67.3  \\
        + & -- & -- & + & 67.4  \\
        + & + & + & -- & 67.5  \\
        + & + & -- & + & 68.1 \\
        + & -- & +  & + & 68.0  \\
        + & + & +  & + & 68.2 \\
         \hline
    \end{tabular}

    \label{tab:fusion_res}
\end{table}

\subsubsection{Final results}

Before evaluating the test set, we trained new models on the combined training and development set, using the parameters that we found optimal in the previous experiments. Simultaneously, we also increased the number of models from 10 to 50, as we saw some improvements by using more DNNs in the ensemble (adding more than 50 models offered little to no improvement on the Dev set). 
Out of the many systems presented before, we selected three to evaluate on the test partition. The first model that we tested was the ensemble trained on the spectral inputs (E2E). Although this system managed to outperform the non-fused SVM systems, it proved to be inferior to the combined models presented in~\cite{compare2020}. Next, we saw that our feature selection method managed to reach markedly higher performance, and it proved to be an equal to the official score in~\cite{compare2020}. Upon inspecting the confidence interval of the feature selection method on the test set (71.3 -- 73.0), we saw that the selection process led to a significantly better model compared to the original ensemble (E2E).  Interestingly, the ensemble using both of these features failed to improve the test set results, suggesting that it might have overfitted for the dev data.

Lastly, we also experimented with the fusion of different models.  
Our aim was to create a more accurate alternative to the on-device ensemble at the expanse of the inference speed. These combined solutions are not applicable to be used in real-time systems, but offer the option of higher accuracy to the users, if they are willing to wait for the results.
We selected the best two ensembles and combined their predictions with those produced by the SVM-based solutions. We found that BoAW and DeepSpectrum SVM models are the most beneficial, so we fused them into our ensemble. The results confirmed that enriching the model types is quite advantageous. The late fusion of the SVMs and the feature-selected E2E ensemble yielded a 77.1 UAR score, which is significantly better than the 71.8 UAR achieved by the fused SVMs.

From the competitions, we selected two works~\cite{Markitantov2020,Szep2020}, which are closely related to ours. In~\cite{Markitantov2020} the authors use an ensemble of ResNet18v2 models trained using various optimization algorithms and cross-validation data folds. Their best result on the MSC 75.9 UAR is better than our purely DNN-based results, which is not surprising since their individual models contain considerably more parameters than ours. On the other hand, fusing with the SVM systems allowed our solution to outperform the purely SVM ensemble markedly. The second work~\cite{Szep2020} employed large pre-trained DNNs and created ensembles by fine-tuning them on the MSC data. This system won the competition by achieving 80.1 UAR. The gap between our best result (77.1) and this 80.1 seems quite big at first glance; however, we should also consider that our entire ensemble of 50 models contained far fewer parameters (4.4 M) than the smallest DNN in their pre-trained ensemble (DenseNet-121 with 8 M). Furthermore, we estimate that their solution is not really applicable in real-life solutions due to the model sizes. Although the authors of~\cite{Szep2020} did not report their inference times, we can estimate it based on the benchmark paper~\cite{Bianco2018Benchmark} analyzing pre-trained Image classification models. Looking at the reported inference times using a Jetson TX1 GPU, we can see that even single pre-trained models would have trouble processing the spectral data in a feasible amount of time. In contrast, our models can easily evaluate data in real-time without using a GPU.

\begin{table*}[!htbp]
    \centering
        \caption{UAR values of our solutions for the Mask Sub-Challenge. The E2E solutions fused 50 models to get the final predictions on the test set.}
        \resizebox{12cm}{2.7cm}{
    \begin{tabular}{c|c|c} \hline
        {\bf System} & {\bf Dev} & {\bf Test}  \\
        \hline
        Fusion of SVM models~\cite{compare2020}& -- & 71.8 \\
        \hline
        E2E & 67.4 & 69.9\\
        E2E selected 10 & 67.9 & \textbf{71.8}\\
        E2E middle fusion (spect + selected 10) & \textbf{68.4} & 70.6\\ \hline
        E2E selected 10 + SVMs & \textbf{71.1} & \textbf{77.1}\\
        E2E middle fusion (spect + selected 10) + SVMs & 70.0 & 75.2\\ \hline 
        E2E ResNet18 ensemble~\cite{Markitantov2020} & -- & 75.9 \\
        Pre-trained E2E ensemble~\cite{Szep2020} & -- & \textbf{80.1} \\
\hline
    \end{tabular}}
    \label{tab:mask_final}
\end{table*}

\subsection{Breathing signal prediction}
On this task, we took a closer look at the E2E baseline system, since that system performs quite well~\cite{compare2020}. The E2E baseline consists of a CNN and a subsequent RNN to make predictions. The CNN is comprised of three stacked one-dimensional CNN layers with 64-128-256 number of filters and 8-6-6 widths of the convolutional layers. Each convolutional layer is followed by a max-pooling layer that subsamples at a stride of 10-8-8. In the baseline system, the convolutional layers are followed by two stacked LSTM-RNN layers with 256 hidden neurons in each RNN layer, which provided the best baseline result \cite{compare2020}. This model outputs a sequence of hidden states, each of which is passed through a linear layer to obtain the breath belt signal prediction.

The above-mentioned system utilizes the raw audio as the input. In our study, as we investigate a gradient-feature selection scheme, so we also experiment with spectral features (like mel-spectrogram) as input for the above system. In our experiments, we used a similar structure to the baseline systems for the raw audio input. The main difference from the baseline architecture is that our model had only one LSTM layer with 100 neurons, followed by a feed-forward layer, which contained 100 rectified linear units before the output layer. In the case of spectral inputs, we found that the best results can be achieved with the architecture that worked best for the MSC. These models contain two CNN layers with 64 neurons and no pooling. After the convolutional part comes the LSTM layer with 100 cells to accumulate the temporal information extracted by the CNNs. Lastly, 100 ReLU units process the LSTM outputs before the output neuron that predicts the belt signal. 

\begin{figure}[t]
    \centering
    \includegraphics[width =\linewidth]{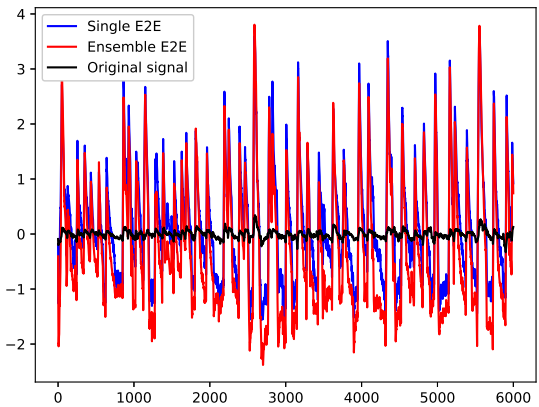}
    \caption{Comparison of the predicted signals using the correlation loss function. Note the difference between the scales of the predictions.}
    \label{fig:corr_mismatch}
\end{figure}

As an initial step, we trained an ensemble system with 10 models using the raw audio input.
The first thing that we noticed is that the models face the issue of mismatch on the scale of the predictions and output labels. Figure~\ref{fig:corr_mismatch}
 depicts the severity of this mismatch. Keep in mind that both the ensemble and its components achieved quite high correlation despite this scaling issue. To alleviate this, we employ a multi-loss scheme using MSE based loss to regularize the baseline correlation loss.

Training an end-to-end system does not have to be restricted to using a single loss function. Often multiple losses are taken into consideration to focus on multiple aspects of the prediction problem. This technique also helps regularize training. 

For breathing sub-challenge, the end-to-end baseline system is trained initially with a correlation-based loss. However, it does not help to bound the outputs to the same scale as the label. To match the output's scale to the label, we use a combination of the correlation loss and the mean squared error (MSE), which can help regularize the end-to-end baseline system. 

\begin{figure}
    \centering
    \includegraphics[width =\linewidth]{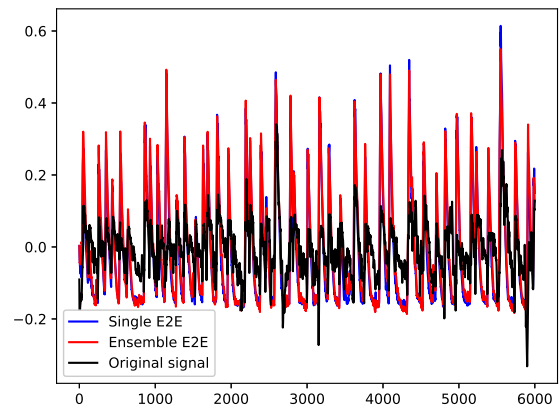}
    \caption{Comparison of the predicted signals using the regularized loss function (corr+MSE). Compare this with Fig.~\ref{fig:corr_mismatch} to see how much smaller the difference of the predictions is (vertical axis).}
    \label{fig:corr_reg_mismatch}
\end{figure}

\begin{table}[!t]
    \centering
        \caption{The table presents the breathing task's Pearson correlation scores for single raw audio-based end-to-end models on the development set (Dev). As we trained ten different models, we present the average and the best result.}
    \begin{tabular}{c|c|c}\hline
        {\bf System} & \multicolumn{2}{c}{{\bf Dev}}\\ \hline
        & Avg. per DNN& Best DNN\\
        \hline
        E2E-corr & .506 & .514\\
        E2E-MSE & .467 & .481 \\
        E2E-corr+MSE & .497 & .521\\ \hline

    \end{tabular}
    \label{tab:breath-ens-vs-single}
    \vspace{-0.5cm}
\end{table}

\begin{table*}[ht]
    \centering
        \caption{The table presents the ensemble E2E model's performance on the breathing task for different loss functions and input features. Inference time calculated as the average time required to evaluate the ensemble of 10 DNNs on one example from the Dev set using a single core of an Intel Xeon W-2133 CPU.}
        \resizebox{12cm}{3cm}{
    \begin{tabular}{c|c|c|c| c}\hline
        \multirow{2}{*}{\bf Input } & \multirow{2}{*}{\bf System (loss func.)}  &\multicolumn{2}{c|}{{\bf Dev}} &{\bf Inf. time}  \\ \cline{3-5}
        & & Corr & MSE & Seconds\\
        \hline
        Audio & single E2E \cite{compare2020} & .507 &  1.682 & 3.5 \\
        \hline
        \multirow{3}{*}{Audio} & E2E-corr & \textbf{.523} & .896 & \multirow{3}{*}{32.6}\\ 
        & E2E-MSE & .480 & .028 & \\
         &E2E-corr+MSE & .514 & .180 & \\ \hline
         \multirow{2}{*}{Spectrum} & E2E-corr & .543  & 1.886 &\multirow{2}{*}{4.4} \\ 
         & E2E-corr+MSE & .540 & .073 & \\ \hline
         \multirow{2}{*}{Selected (output)} & E2E-corr & .539 & 1.773 & \multirow{4}{*}{3.4}\\ 
         & E2E-corr+MSE & .517 & .086 & \\ \cline{1-4}
        \multirow{2}{*}{Selected (loss)} & E2E-corr & .535 & 1.293 &  \\
         & E2E-corr+MSE & .510 & .097 &  \\ \hline
    \end{tabular}}
    \label{tab:breath-dev}
    \vspace{-0.1cm}
\end{table*}

\begin{table*}[!bp]
    \centering
        \caption{Comparison of ensemble E2E model's performance on the test set of breathing task.}
                \resizebox{12cm}{3cm}{
    \begin{tabular}{c|c| c}\hline
        {\bf Input } & {\bf System (loss func.)}  & {\bf Test (Corr)}  \\
        \hline
        Audio & Baseline (E2E) \cite{compare2020} & .731 \\
        \hline
        \multirow{2}{*}{Audio} & E2E-corr & \textbf{.759}\\
         &E2E-corr+MSE  & .751\\ \hline
         \multirow{2}{*}{Spectrum} & E2E-corr & .740 \\
         & E2E-corr+MSE  & .737\\ \hline
         \multirow{2}{*}{Selected (output)} & E2E-corr  & .721 \\
         & E2E-corr+MSE & .700\\ \hline
        \multirow{2}{*}{Selected (loss)} & E2E-corr & .704 \\
         & E2E-corr+MSE & .704 \\ \hline
         Audio+Spectrum & ResNet18v2 ensemble \cite{Markitantov2020} & \textbf{.763} \\ \hline
    \end{tabular}}
    \label{tab:breath}

\end{table*}

Table~\ref{tab:breath-dev} summarizes the effects of using different loss functions during training. Even though the baseline system produced high correlation values, it had the highest MSE value. Combining the predictions of 10 models (corr) reduced the MSE significantly and outperformed the baseline results. Using the MSE as loss function performed the worst but produced the lowest MSE. Lastly, we can see that using the multi-loss ensemble of E2E model (E2E-corr+MSE) slightly lowers the achieved correlation; on the other hand, in terms of MSE, it is much better than E2E-corr. Figure~\ref{fig:corr_reg_mismatch} depicts how well the combined loss regularizes the network. We can see that penalizing the MSE during training helps the DNNs to predict the signal with high correlation and on the correct scale.

Next, we inspected the performance of the individual DNNs in the ensembles (see Table~\ref{tab:breath-ens-vs-single}). We once again saw that depending on the random initialization, the networks achieved varying performance. Furthermore, even the best models were inferior to the ensemble.

Our next set of experiments focused on reducing the model size and its inference time. We achieved this by using spectral features and our saliency map-based feature selection. For simplicity's sake, we selected the 10 most important features, just like in the case of MSC. Table~\ref{tab:breath-dev} shows the results and processing speed of models using different inputs. The first thing to notice is the considerable reduction in inference time once we switch to spectral inputs. This vast difference is mainly thanks to the reduced amount of data that the DNNs have to process. Interestingly, using spectral inputs also helps the model to achieve better performance on the development data. Using the 10 selected frequency bands led to a slight drop in performance, but it also reduced the processing time by ~25\%. Considering that each example had a fixed length of 4 minutes, we can safely say that even all of our ensembles can process data in real-time.

We tried using the middle fusion approach to combine different input modalities, but it did not help in this case. The models using more than one input type proved to be equal or worse than those that relied on the spectrum or raw audio.

In Table~\ref{tab:breath}, we present the results of 10-model ensembles to compare with the baseline performance. E2E-corr+MSE was slightly worse on the evaluation set than the E2E-corr ensembles in terms of overall correlation. Furthermore, the raw audio input proved to be the best feature set, slightly outperforming the spectral models. The best ensemble of E2E corr. managed to outperform the baseline result and showed an absolute improvement of 2.8 correlation points over the baseline result. We can also observe that reducing the input has slightly degraded the performance, but keep in mind that these smaller models became significantly faster.

We also compared our best solutions with the official winner's system~\cite{Markitantov2020}. We can see that their ensemble got a slightly better correlation value on the evaluation data, but the difference is minimal. Two factors could be listed as the reason for their better performance. First, their models were significantly larger than ours containing approximately 2M trainable parameters, while our spectral DNNs had 167k, which was further reduced to 106k after feature selection. Secondly, \cite{Markitantov2020} used a cross-validation training scheme, meaning that more data was used to train the DNNs. We decided to avoid utilizing the Dev set for training since several instances in that set were problematic (the models produced low correlation for these files). Overall, we can say that our model achieved comparable results at a much faster pace.


\section{Conclusion}

In this work, we employed AI-enabled tools to tackle two computational paralinguistic tasks, which process the speaker's audio to potentially, help monitoring and diagnosing COVID-19 related conditions. To effectively reach the general public, these systems must be run on mobile devices; however, using AI tools on hand-held devices poses constraints to these methods' resource utilization. Our work focused on creating models that can meet these requirements while reaching high performance. We presented a novel ensemble neural network-based feature selection method, which can help us reduce the size of the model and consequently, its memory and processing power consumption.

On the mask detection task, our feature selection method provided improved performance over the baseline method while reducing the on-device footprint by a $10^{th}$ and reducing the inference time by a third. In combination with other simple approaches, our solution was able to outperform a far larger ResNet-based ensemble and close the gap with the state-of-the-art. 

On a breathing state detection task, we achieved competitive performance to baselines while reducing the on-device footprint by 36\% and inferring the result 25\% faster. Compared to the baseline solution, our method took only one-eighth of the time to produce the output while reaching comparable correlation values.
Our proposed method could enable smartphone applications to reach deployment quickly and efficiently. 


\bibliographystyle{IEEEtran}
\bibliography{compare.bib,IS2020.bib}

%

\begin{IEEEbiography}[{\includegraphics[width=1in,height=1.25in,clip,keepaspectratio]{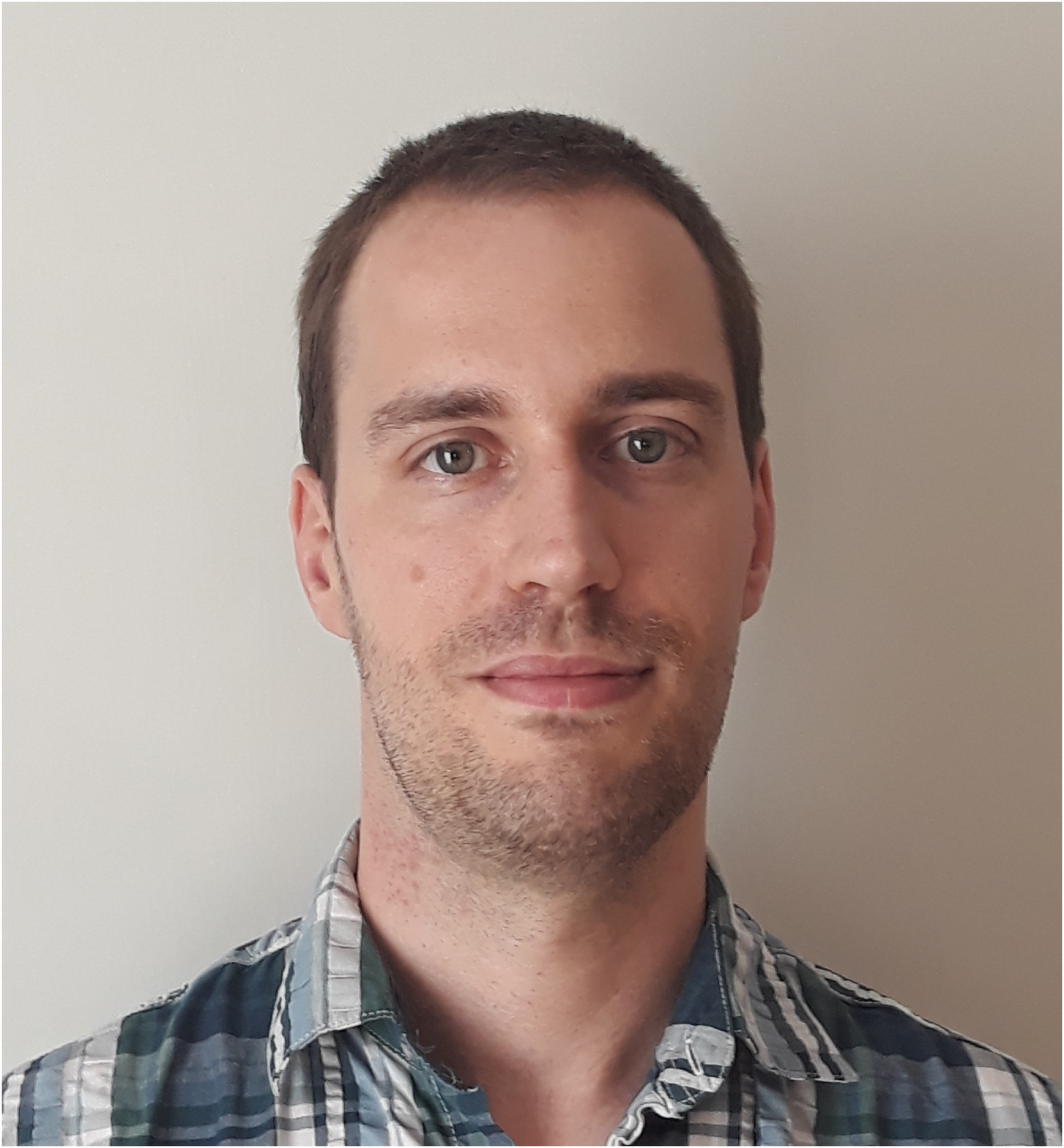}}]{Tamás~Grósz}
received a PhD degree in speech recognition from the University of Szeged, in 2018. Between 2017 and 2018, he worked as an assistant research fellow at the  Hungarian Academy of Sciences' Research Group on Artificial Intelligence. From 2018 till 2019, he was a senior lecturer at the Department of Computer Algorithms and Artificial Intelligence, University of Szeged. He is currently a Research Fellow at the Department of Signal Processing and Acoustics, Aalto University. His current research focuses on automatic speech recognition, deep learning, and computational paralinguistics.
\end{IEEEbiography}

\begin{IEEEbiography}[{\includegraphics[height=1.15in,trim={150 90 250 0},clip]{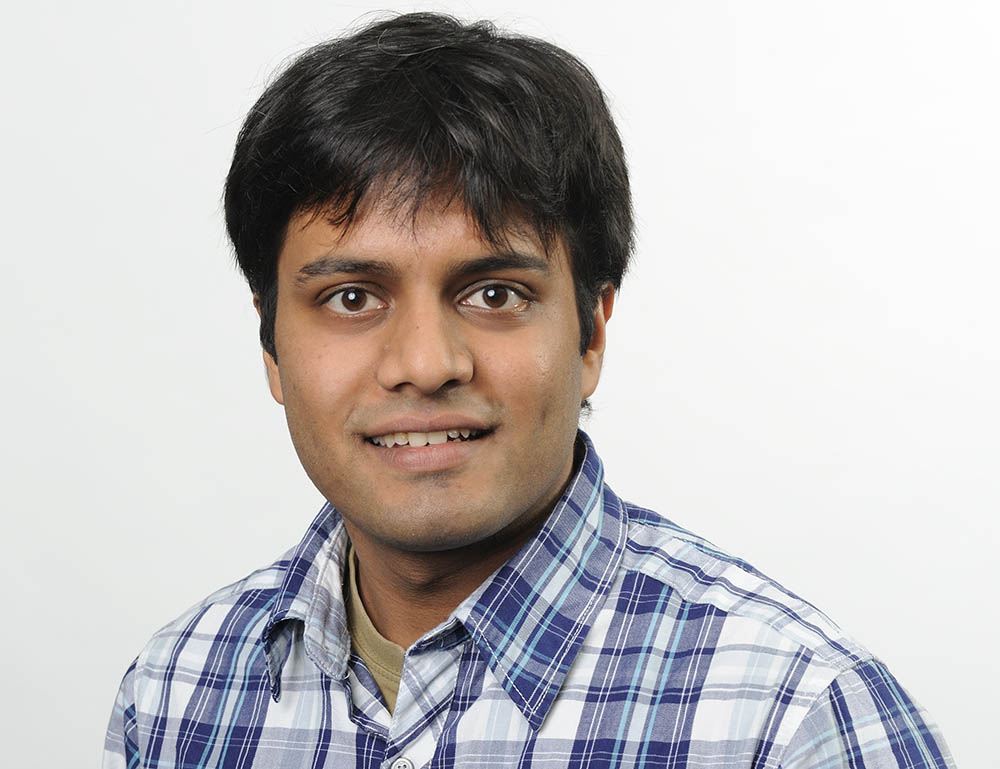}}]{Mittul~Singh} is specialized in building speech recognition based solutions and he has a PhD in Computer Science from Saarland University in Germany. He is currenttly working as Senior AI scientist at an AI startup SiloAI. Before joining Silo AI, Mittul worked as a postdoctoral researcher at the Speech Recognition Group at Aalto University. Mittul is passionate about developing deep learning solutions for restricted data scenarios to reach new markets and customers.
\end{IEEEbiography}


\begin{IEEEbiography}[{\includegraphics[width=1.3in,height=1.35in,clip,keepaspectratio]{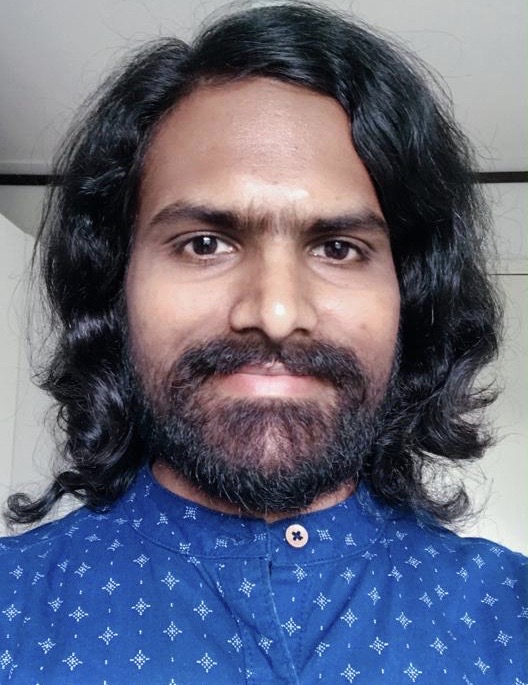}}]{Sudarsana~Reddy~Kadiri} received Bachelor of Technology degree from Jawaharlal Nehru Technological University (JNTU), Hyderabad, India, in 2011, with a specialization in Electronics and Communication Engineering (ECE), the M.S. (Research) during 2011-2014, and later converted to Ph.D., and received Ph.D. degree from the Department of ECE, International Institute of Information Technology, Hyderabad (IIIT-H), India in 2018. He was a Teaching Assistant for several courses at IIIT-H during 2012-2018. He is currently a Postdoctoral Researcher with the Department of Signal Processing and Acoustics, Aalto University, Espoo, Finland. His research interests include signal processing, speech analysis, speech synthesis, paralinguistics, affective computing, voice pathologies, machine learning and auditory neuroscience.
\end{IEEEbiography}

\begin{IEEEbiography}[{\includegraphics[width=1.1in,height=1.3in,clip,keepaspectratio]{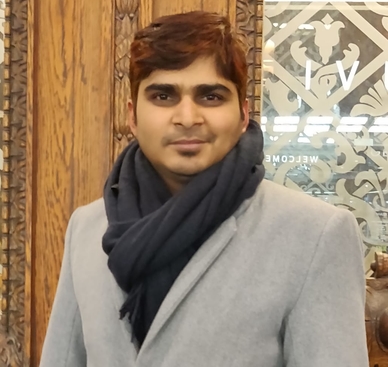}}]{Hemant~Kathania} received his M.Tech. degree in Electronics and Electrical Engineering from the Indian Institute of Technology Guwahati, India, in 2012. He then obtained his Ph.D. degree from the Department of Electronics and Communication Engineering, National Institute of Technology Sikkim in 2018, He worked as an Assistant Professor in the Department of Electronics and Communication Engineering, National Institute of Technology (NIT) Sikkim, India from 2013 to 2019. He is currently working as an Postdoctoral Researcher in the Department of Signal Processing and Acoustics, Aalto University, Finland. He is a member of IEEE and IEE signal processing society. His current research interests include speech signal processing, speech recognition, keyword spotting and speaker verification.

\end{IEEEbiography}

\begin{IEEEbiography}[{\includegraphics[width=1in,height=1.25in,clip,keepaspectratio]{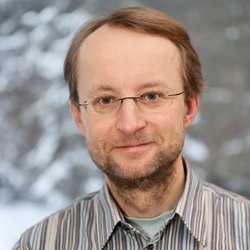}}]{Mikko Kurimo} 
received the D.Sc. (Ph.D.) in technology degree in computer science from Helsinki University of Technology, Espoo, Finland, in 1997. He is currently a Full Professor in the Department of Signal Processing and Acoustics, Aalto University. His research interests include speech recognition, machine learning, and natural language processing.
\end{IEEEbiography}

\EOD

\end{document}